\documentclass[conference,a4paper]{IEEEtran}

\usepackage{cite}      

\usepackage{graphicx}  

\usepackage{psfrag}    

\usepackage{amsmath}   
\usepackage[absolute]{textpos}
\newcommand{\copyrightstatement}{
    \begin{textblock}{15}(0.5,0.3)    
         \noindent
         \centering
         \textblockcolour{white}
         \footnotesize
         \copyright 2013 IEEE. Personal use of this material is permitted. Permission from IEEE must be obtained for all other uses, in any current or future media, including reprinting/republishing this material for advertising or promotional purposes, creating new collective works, for resale or redistribution to servers or lists, or reuse of any copyrighted component of this work in other works
    \end{textblock}
}

\begin{document}

\title{Modeling the Energy Consumption \\ of HEVC Intra Decoding \\ \vspace{0.15cm} \large{\it{Invited Paper}}}

\copyrightstatement



%
\author{\authorblockN{Christian Herglotz, Dominic Springer, Andrea Eichenseer, 
and Andr\'e Kaup\vspace{0.15cm}}
\authorblockA{ Multimedia Communications and Signal Processing\\
Friedrich-Alexander-University Erlangen-Nuremberg,
Cauerstr. 7, 91058 Erlangen, Germany\\ Email: \{ \ herglotz, springer, eichenseer, kaup\ \}\ @LNT.de}\vspace{-0.3cm}}


\maketitle

\begin{abstract}
Battery life is one of the major limitations to mobile device use, which makes research on energy efficient soft- and hardware an important task. This paper investigates the energy required by a CPU when decoding compressed bitstream videos on mobile platforms. A model is derived that describes the energy consumption of the new HEVC decoder for intra coded videos. We show that the relative estimation error of the model is smaller than $3.2\%$ and that the model can be used to build encoders aiming at minimizing decoding energy. 
\end{abstract}


%
\IEEEpeerreviewmaketitle

\section{Introduction}
\label{sec:int}


 
In recent years, the evolution of bandwidth for mobile communication and the development of smart and powerful portable devices has led to a tremendous increase in online and live video streaming. Studies indicate that nowadays over 50\% of internet traffic is based on video content \cite{cisco}, the decoding of which has to be performed on the displaying device. 
For portable devices that receive their power from batteries, this decoding process is one of the most demanding energy drains and can reduce the available battery operating time significantly. Therefore, the search for energy-efficient video decoding techniques is a valuable task in the pursuit of extended operating time for portable devices. 

Recently, a new video coding scheme, High Efficiency Video Coding (HEVC) \cite{Sullivan12}, was released as the successor to the heretofore predominant H.264\slash AVC video coding standard. 
In this paper, the intra mode of this new standard is investigated with respect to its energy consumption. The results show that different intra coding techniques 
consume different amounts of energy. This information can, e.g., be exploited to enable energy-aware video coding. 

For H.264\slash AVC several complexity and energy models for the decoder have been proposed \cite{Lee07b, Li12}. 
In this paper, such a model is presented for HEVC, for which, to the knowledge of the authors, no model has been developed yet.

The following section describes the test setup used to measure the energy consumption on a portable-like platform. Section \ref{sec:HEVC} summarizes the HEVC intra coding mode and Section \ref{sec:method} presents how the energy consumption of different decoding modules was investigated. The resulting model is formally presented in Section \ref{sec:decModel}. Finally, Section \ref{sec:eval} shows the validity of the model by comparing the energy estimations with actual measurements. Section  \ref{sec:concl} concludes this paper.



\section{Test Setup}
\label{sec:setup}

\begin{figure}
	\centering
	\psfrag{V}[l][c]{$V_\mathrm{0}$}
	\psfrag{M}[c][c]{$v_\mathrm{A}(t)$}
	\psfrag{I}[c][c]{$i(t)$}
	\psfrag{P}[r][c]{$v_\mathrm{DUT}(t)$}
	\psfrag{A}[c][c]{A}
	\psfrag{D}[c][c]{DUT}
	\includegraphics[width=0.35\textwidth]{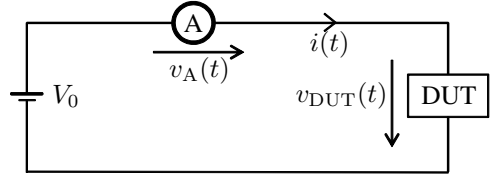}
	\caption{Test setup with voltage source $V_\mathrm{0}$, current $i(t)$, voltage across ampere meter $v_\mathrm{A}(t)$, and voltage across the decoding device $v_\mathrm{DUT}(t)$. }
	\label{fig:measSetup}
	\vspace{-0.5cm}
\end{figure}

A block diagram of the test setup is depicted in Figure \ref{fig:measSetup}. As a voltage source, the HAMEG HM7042 with an output voltage $V_\mathrm{0}=5.2\mathrm{V}$ is used. The specifications indicate that the maximum voltage drift during operation is smaller than $2.5\mathrm{mV}$. Hence, we assume the voltage to be constant for the conducted measurements. The current $i(t)$ is measured using the Agilent 34401A Multimeter. 
The device under test (DUT) on which the decoder is executed is a PandaBoard \cite{Panda} featuring an OMAP4430 System-on-Chip (SoC). This SoC holds two 1GHz Cortex-A9 ARM processors and is frequently used on portable devices from well-known manufacturers like LG \cite{LG2} and Amazon \cite{amazon2}. Ubuntu 12.04 was chosen as the operating system. Running the decoder on runlevel 1 with only a few basic modules loaded assures that the measurements are not disturbed by running background services. 
Reproducible results are generated by setting the decoding process to the highest priority and executing it on a single, predefined core. Onboard indicators like LEDs are switched off. Commands are sent via a serial COM-Port where it was ensured that energy consumption from this port does not affect the measurements. 

The reference software HM 9.1 \cite{HM-9.1} is used as encoding and decoding software. Furthermore, to suppress energy consuming access to the flash storage, the executables and the encoded videos are stored on a ramdisk, and the output of the decoder (the decoded video) is not saved. 
The code was compiled using GCC 4.6.3 with -O3 optimization. 

The energy of one single decoding process $E_\mathrm{dec}$ is calculated as
\begin{equation}
	E_\mathrm{dec} = E_\mathrm{all} - E_\mathrm{idle}, 
\label{eq:energy}
\end{equation}
where $E_\mathrm{idle}$ is the energy the DUT consumes in idle mode (no user process active). $E_\mathrm{all}$ is the complete energy consumed during decoder execution (including idle energy). The difference $E_\mathrm{dec}$ finally yields the energy solely required by the decoder. $E_\mathrm{all}$ and $E_\mathrm{idle}$ are obtained by
\begin{align}
	E & = \int_0^{T_\mathrm{int}}{v_\mathrm{DUT}(t)\cdot i(t)\, \mathrm{d}t} = \int_0^{T_\mathrm{int}}{\left(V_\mathrm{0}-v_\mathrm{A}(t)\right) \cdot i(t)\, \mathrm{d}t} \notag \\   
	& = \int_0^{T_\mathrm{int}}{\left(V_\mathrm{0}-i(t)\cdot R_\mathrm{A}\right) \cdot i(t)\, \mathrm{d}t} \notag \\
	& = V_\mathrm{0}\int_0^{T_\mathrm{int}}{i(t)\, \mathrm{d}t} - R_\mathrm{A}\int_0^{T_\mathrm{int}}{i^2(t)\, \mathrm{d}t} \notag \\
	&	\approx V_\mathrm{0}\cdot I_\mathrm{A}\cdot T_\mathrm{int} - R_\mathrm{A}\cdot I^2_\mathrm{A}\cdot T_\mathrm{int}\, , \label{eq:int}
\end{align}
where $R_\mathrm{A}=0.1\mathrm{\Omega}$ is the shunt of the ampere meter and $I_\mathrm{A}$ is the current measured with an integration time $T_\mathrm{int} = 2\mathrm{s}$. The error resulting from the approximation is smaller than the standard deviation of the measurements and is thus ignored. 

\begin{figure}
	\centering
	\psfrag{012}[c][c]{\footnotesize{Time [s]}}
	\psfrag{011}[bc][c]{\footnotesize{Current [A]}}
	\psfrag{000}[c][c]{\footnotesize{$0$}}
	\psfrag{001}[c][c]{\footnotesize{$4$}}
	\psfrag{002}[c][c]{\footnotesize{$8$}}
	\psfrag{003}[c][c]{\footnotesize{$12$}}
	\psfrag{004}[r][r]{\footnotesize{$0.46$}}
	\psfrag{005}[r][r]{\footnotesize{$0.5$}}
	\psfrag{006}[r][r]{\footnotesize{$0.54$}}
	\psfrag{007}[r][r]{\footnotesize{$0.58$}}
	\psfrag{008}[r][r]{\footnotesize{$0.62$}}
	\psfrag{009}[l][l]{\footnotesize{Decode}}
	\psfrag{010}[l][l]{\footnotesize{Idle}}
	\includegraphics[width=0.5\textwidth]{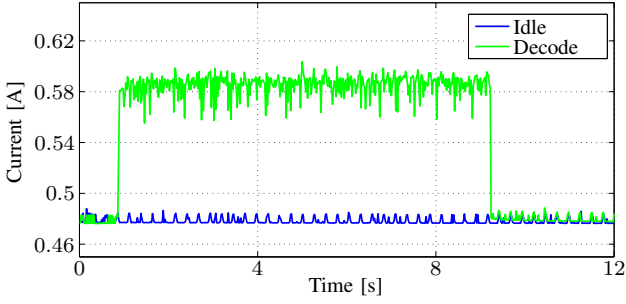}
	\vspace{-0.8cm}
	\caption{Current during decoding and idle mode. In idle mode, no user processes are active. The decoded video consists of 17 intra frames with a resolution of $416\times 240$ pixels. }
	\label{fig:0.02}
	\vspace{-0.5cm}
\end{figure}

Figure \ref{fig:0.02} illustrates the characteristics of the current while decoding 17 frames of a $416\times 240$ pixel, intra coded video. Integration time is set to $4\mathrm{ms}$, where one measurement is performed every $20\mathrm{ms}$. 
The area in between both curves multiplied by the voltage describes the energy that is consumed by the decoder. During the decoding process, the variation of the current is faster than the sampling rate of the ampere meter, thus we decided to set the integration time $T_\mathrm{int}$ to $2\mathrm{s}$. The decoding process occurs during these two seconds, assuring that the complete process is covered. Accordingly, only one value is returned for each measurement.

\section{Intra Coding in the HEVC Standard}
\label{sec:HEVC}
This section provides a short introduction to the intra coding mode of HEVC. A detailed description can be found in \cite{Sullivan12} and \cite{Lainema12}. 

Each frame is decomposed into coding tree units (CTUs) with a size of $64\times 64$ pixels. These CTUs are further split into rectangular coding units (CUs) forming a quadtree decomposition. The depth describes how often the CTU is recursively split into CUs; the maximum depth is 3 with a CU size of $8\times 8$ pixels. A CU further consists of a prediction unit (PU) and a transform unit (TU). 
The PU provides the information about the intra prediction mode (planar, DC, or angular). Depending on the neighborhood, this mode can be coded as a so-called most probable mode (MPM). The TU holds transform coefficients for the compensation of the prediction error. 
The terms coding block (CB), prediction block (PB), and transform block (TB) refer to one color component (Y, Cb, Cr) of the respective units. PBs and TBs may have a minimum size of $4\times 4$ pixels. 

For each TB, a coded block flag (CBF) is set signaling the existence of nonzero coefficients. If the CBF is `true', 
the coefficients are coded using a significance map that indicates the positions of the nonzero coefficients. Subsequently, the respective values are coded successively using a Golomb-Rice code. 

It is worth mentioning that in intra mode, the actual prediction is always performed at the same depth as the transformation of the coefficients. Furthermore, a flag signals whether the transformation is skipped (transform skip flag TSF). This flag only exists for TBs of size $4\times 4$. 

To sum up, the following list shows the coding parameters that were considered for this work:
\begin{itemize}
\item depth, 
\item prediction mode (planar, DC, or angular), 
\item prediction mode coding, 
\item number of residual coefficients, 
\item absolute value of residual coefficients, 
\item transform skipping. 
\end{itemize}

\section{Measuring the Decoding Energy}
\label{sec:method}
\begin{figure}
	\centering
	\psfrag{a}[c][c]{$E_\mathrm{test}$}
	\psfrag{b}[c][c]{$E_\mathrm{ref}$}
	\psfrag{c}[c][c]{$E_\mathrm{rem}$}
	\includegraphics[width=0.35\textwidth]{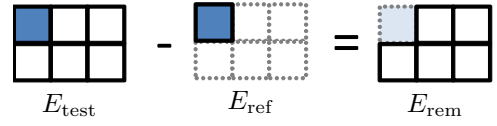}
	\vspace{-0.5cm}
	\caption{Calculation method for deriving the energy consumption of five CTUs. Each square depicts one CTU, where the first (blue) CTU depicts the defined CTU. The remaining (white) CTUs are coded using constant parameters. The dashed squares are depicted for clarity reasons, they are not part of the respective measured energy. }
	\vspace{-0.5cm}
	\label{fig:CUenergy}
\end{figure}

To obtain energies dependent on the coding parameters, we coded test videos containing one single frame ($384\times 192$ pixels which corresponds to 18 CTUs). In addition, a single-CTU reference video was coded, where we defined the single CTU to be equal to the first CTU in the test videos. Subtracting the measured energy of the reference video $E_\mathrm{ref}$ from the energy of the test video $E_\mathrm{test}$  yields the energy required to decode the remaining 17 CTUs $E_\mathrm{rem}$. When these CTUs are coded using the same coding parameters for each sub-CU, 
dividing $E_\mathrm{rem}$ by the number of sub-CUs returns the energy required to decode one single sub-CU. Figure \ref{fig:CUenergy} visualizes this calculation method for the case of six CTUs.

During initial measurements, basic energies needed for startup and termination of the decoding process were deduced. These consist of a constant energy offset and the energy required for the decoding of slice parameters. Dedicated measurements yielded an offset energy of $E_\mathrm{0} = 1.579\cdot 10^{-2}\mathrm{J}$ and a slice energy of $e_\mathrm{slice} = 6.250\cdot 10^{-4}\mathrm{J}$. 

\begin{figure}
	\centering
	\psfrag{000}[c][c]{\footnotesize{Pla}}
	\psfrag{001}[c][c]{\footnotesize{DC}}
	\psfrag{002}[c][c]{\footnotesize{Ver}}
	\psfrag{003}[c][c]{}
	\psfrag{004}[c][c]{}
	\psfrag{005}[c][c]{}
	\psfrag{006}[c][c]{\footnotesize{Ang}}
	\psfrag{007}[c][c]{}
	\psfrag{008}[c][c]{}
	\psfrag{009}[c][c]{}
	\psfrag{010}[c][c]{\footnotesize{Hor}}
	\psfrag{011}[r][r]{\footnotesize{$0.02$}}
	\psfrag{012}[r][r]{\footnotesize{$0.03$}}
	\psfrag{013}[l][l]{\footnotesize{depth 4}}
	\psfrag{014}[l][l]{\footnotesize{depth 3}}
	\psfrag{015}[l][l]{\footnotesize{depth 2}}
	\psfrag{016}[l][l]{\footnotesize{depth 1}}
	\psfrag{017}[c][b]{\footnotesize{Intra mode}}
	\psfrag{018}[bc][bc]{\footnotesize{Consumed energy [J]}}
	\includegraphics[width=0.5\textwidth]{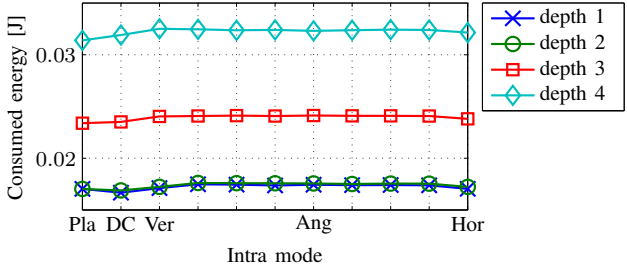}
	\caption{Consumed energy for 17 CTUs depending on depth and intra mode (intra modes 0 to 10). Transformation coefficients and CBFs are set to `zero' and `false', respectively. }
	\label{fig:depth_modes}
\end{figure}

In the second test series, only the energy consumption of the prediction was measured. Figure \ref{fig:depth_modes} shows the decoding energy of 17 CTUs depending on intra mode and depth. In the test videos, the CBFs are set to zero (which results in zero-valued coefficients). However, it has to be noted that the transformation of the (zero-valued) residuals is performed anyhow as the reference software provides no means to skip it in intra mode (except when utilizing the TSF). 
To maintain clearity, only the results of the first ten intra modes are shown, since measurements proved that the energy consumption for the remaining modes is equal to the depicted modes. 
From this diagram, $4\times 4$ mode dependent energies were extracted, one for each depth and for the following four prediction modes: 
\begin{itemize}
\item planar (pla; mode 0), 
\item DC (dc; mode 1),
\item horizontal, vertical, and diagonal (hvd; modes 2, 10, 18, 26, and 34), 
\item angular (ang; remaining modes). 
\end{itemize}
The measured energies required for decoding one single PU depending on depth and coding mode are shown in Table \ref{tab:depthnMode} and will be referred to as $e_\mathrm{mode, depth}$. Additionally, the last row shows the mean energies across the intra modes (avg). 
\begin{table}
\renewcommand{\arraystretch}{1.3}
\vspace{-0.4cm}
\caption{Mean values of decoding energy $e_\mathrm{mode, depth}$ per PU [J] }
\vspace{-0.7cm}
\label{tab:depthnMode}
\begin{center}
\begin{tabular}{|l||c||c||c||c|}
\hline
& depth 1 & depth 2 & depth 3 & depth 4\\
\hline\hline
 pla & $2.505\cdot 10^{-4}$ & $6.262\cdot 10^{-5}$ & $2.150\cdot 10^{-5}$ & $7.214\cdot 10^{-6}$\\
\hline
dc & $2.452\cdot 10^{-4}$ & $6.209\cdot 10^{-5}$ & $2.161\cdot 10^{-5}$ & $7.332\cdot 10^{-6}$\\
\hline
hvd & $2.512\cdot 10^{-4}$ & $6.336\cdot 10^{-5}$ & $2.199\cdot 10^{-5}$ & $7.429\cdot 10^{-6}$\\
\hline
ang & $2.556\cdot 10^{-4}$ & $6.442\cdot 10^{-5}$ & $2.215\cdot 10^{-5}$ & $7.443\cdot 10^{-6}$ \\
\hline
avg & $2.550\cdot 10^{-4}$ & $6.262\cdot 10^{-5}$ & $2.150\cdot 10^{-5}$ & $7.431\cdot 10^{-6}$ \\
\hline
\end{tabular}
\end{center}
\vspace{-0.8cm}
\end{table}

The third test series considers the transformation of the residual coefficients. For TBs of size $4\times 4$, the amount of energy saved by skipping the transformation process using TSFs was measured. 
The resulting value adds up to $e_\mathrm{TSF}=5.0916\cdot 10^{-7}\mathrm{J}$. 

Furthermore, decoding an intra mode that is not one of the MPMs requires additional energy. This is accounted for by introducing the factor $e_\mathrm{noMPM} = 7.413\cdot 10^{-7}\mathrm{J}$. 

Investigating the energy required for decoding residual coefficients, measurements showed that it is reasonable to split up the process into three parts: CBFs, number of coefficients, and value of coefficients.  

The energy depending on the number of CBFs comprises the startup of the coefficient decoding process of a single TB. As reference videos, the coded test video files from the first test series were used, where all coefficients were set to `zero'. In the CBF test files, the CBFs were set to `true', and the first coefficient of each TB was set to `one'. The difference between these two decoding energies was used to determine the consumed energy. The mean value was found to be $e_\mathrm{CBF}=9.863\cdot 10^{-7}\mathrm{J}$ per set CBF.

\begin{figure}
	\centering
	\psfrag{000}[c][c]{\footnotesize{$0$}}
	\psfrag{001}[c][c]{\footnotesize{$256$}}
	\psfrag{002}[c][c]{\footnotesize{$512$}}
	\psfrag{003}[c][c]{\footnotesize{$768$}}
	\psfrag{004}[c][c]{\footnotesize{$1024$}}
	\psfrag{005}[r][r]{\footnotesize{$0.02$}}
	\psfrag{006}[r][r]{\footnotesize{$0.025$}}
	\psfrag{007}[r][r]{\footnotesize{$0.03$}}
	\psfrag{008}[l][l]{\footnotesize{Approximation}}
	\psfrag{009}[l][l]{\footnotesize{Measurement}}
	\psfrag{010}[tc][tc]{\footnotesize{Number of one-valued coefficients}}
	\psfrag{011}[bc][bc]{\footnotesize{Consumed energy [J]}}
	\includegraphics[width=0.45\textwidth]{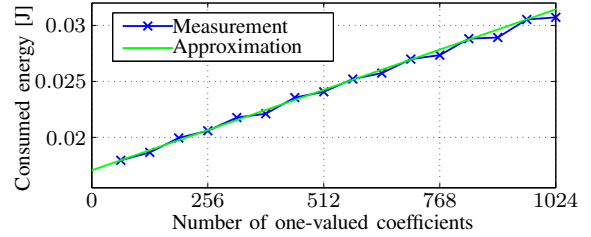}
	\vspace{-0.2cm}
	\caption{Measured and approximated consumed energy depending on number of non-zero luma coefficients per sub-CU. CTU depth is 1, chroma coefficients are zero. }
	\label{fig:nrCoeffs}
\end{figure}

To deduce the energy required to decode a single coefficient, a predefined number of one-valued coefficients was coded. 
Figure \ref{fig:nrCoeffs} shows the measured energy depending on the number of one-valued coefficients inside the CTUs. The curve is approximated linearly with an energy of $e_\mathrm{coeff}=2.064\cdot 10^{-7}\mathrm{J}$ per coefficient. 

\begin{figure}
	\centering
	\psfrag{000}[c][c]{\footnotesize{$10^0$}}
	\psfrag{001}[c][c]{\footnotesize{$10^1$}}
	\psfrag{002}[c][c]{\footnotesize{$10^2$}}
	\psfrag{003}[c][c]{\footnotesize{$10^3$}}
	\psfrag{004}[c][c]{\footnotesize{$10^4$}}
	\psfrag{005}[r][r]{\footnotesize{$0.04$}}
	\psfrag{006}[r][r]{\footnotesize{$0.045$}}
	\psfrag{007}[r][r]{\footnotesize{$0.05$}}
	\psfrag{008}[r][r]{\footnotesize{$0.055$}}
	\psfrag{009}[l][l]{\footnotesize{Approximation}}
	\psfrag{010}[l][l]{\footnotesize{Negative values}}
	\psfrag{011}[l][l]{\footnotesize{Positive values}}
	\psfrag{012}[tc][tc]{\footnotesize{Coefficient value $c$}}
	\psfrag{013}[bc][bc]{\footnotesize{Consumed energy [J]}}
	\vspace{-0.3cm}
	\includegraphics[width=0.5\textwidth]{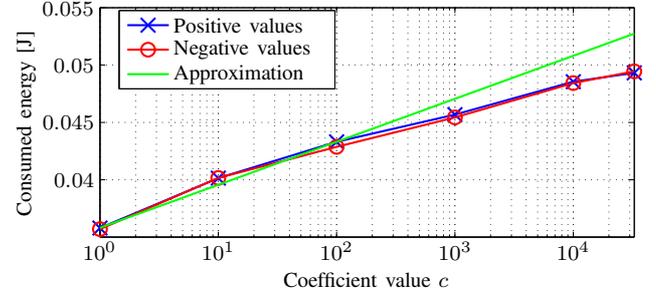}
	\vspace{-0.5cm}
	\caption{Measured and approximated decoding energy with respect to the coefficient value (one non-zero value for each color component, depth 4). Measurements are made for positive (blue) and negative (red) coefficients; the approximation only considers the absolute value. }
	\label{fig:valCoeffs}
	\vspace{-0.5cm}
\end{figure}

In the final test series, only one coefficient was coded for every TB, representing a constant positive or negative value $c$. 
Figure \ref{fig:valCoeffs} depicts the energy required to decode coefficients depending on their absolute value. The measurements show that the relation is close to logarithmic, which corresponds to the complexity properties of the employed Golomb-Rice code \cite{Sole12}. The curve is approximated by multiplying the sum of the logarithms to the base $2$ of the values with the constant factor $e_\mathrm{val} = 1.729\cdot 10^{-7}\mathrm{J}$. $e_\mathrm{val}$ is chosen to fit best for the more frequently occuring coefficient values smaller than $256$.

\section{Decoder Energy Model}
\label{sec:decModel}
\subsection{Accurate Model}

Closely investigating the measured data presented in Section \ref{sec:method} a model to estimate the decoding energy $E_\mathrm{dec}$ was deduced. The respective estimated energy $\hat E_\mathrm{dec}$ is described by Equation (\ref{eq:eDecode}). 
The values and meaning of the energy constants have already been explained in Section \ref{sec:method}. 
The variable $n$ denotes quantities, where $n_\mathrm{slice}$ is the number of slices, $n_\mathrm{mode, depth}$ the number of TUs using the respective intra mode at the respective depth, $n_\mathrm{CBF}$ the number of coded block flags, $n_\mathrm{coeff}$ the total amount of nonzero coefficients, $n_\mathrm{noMPM}$ the number of intra modes that were not coded as MPMs,
 and $n_\mathrm{TSF}$ the number of TSFs. 
 The latter are subtracted as the transformation energy is included in $e_\mathrm{mode,depth}$.  
\begin{align}
  \hat E_\mathrm{dec}  = \ & E_0 + e_\mathrm{slice}\cdot n_\mathrm{slice} \notag \\  
& +  \sum_\mathrm{depth=1}^4{\left(\sum_\mathrm{mode=0}^{34}{e_\mathrm{mode, depth}}\cdot n_\mathrm{mode, depth}\right)} \notag \\ 
& +  e_\mathrm{CBF}\cdot n_\mathrm{CBF} + e_\mathrm{coeff}\cdot n_\mathrm{coeff}   \notag  \\
& +  e_\mathrm{val}\cdot \sum_{\forall c\ne 0}{\log _2 \left| c\right|} \notag \\
& + e_\mathrm{noMPM}\cdot n_\mathrm{noMPM} -  e_\mathrm{TSF}\cdot n_\mathrm{TSF}. \label{eq:eDecode}
\end{align}



\subsection{Simplified Model}
Evaluating the accurate model we found that some of the parameters only contribute marginally to the consumed energy. First, we could see that intra mode coding and transform skipping can be neglected without significant impact. Second, merging the four intra mode constants into one mean constant (cf. Table \ref{tab:depthnMode}) only leads to minor impairment as the separate mode energies differ by less than $4\%$. Third, the coefficient values can be disregarded if the quantization parameter (QP) is greater than 30. Equation (\ref{eq:eDecode}) thus reduces to 
\begin{align}
  \hat E_\mathrm{dec}  = \ & E_0 + e_\mathrm{slice}\cdot n_\mathrm{slice} \notag \\  
& +  \sum_\mathrm{depth=1}^4{e_\mathrm{depth}\cdot n_\mathrm{depth}} \notag \\ 
& +  e_\mathrm{CBF}\cdot n_\mathrm{CBF} + e_\mathrm{coeff}\cdot n_\mathrm{coeff} \, ,   \label{eq:eSimpleDecode}
\end{align}
where $e_\mathrm{depth}$ is the mean energy across all intra modes and $n_\mathrm{depth}$ is the sum of all intra coded CUs in the respective depth. 

Both models are constructed in such a way that the complete decoding energy can easily be estimated on the basis of encoding information. By extending the encoder through dedicated counters, the energy can immediately be calculated during the encoding process.

\section{Model Evaluation}
\label{sec:eval}

\begin{figure}
	\centering
	\psfrag{000}[c][c]{\footnotesize{1}}
	\psfrag{001}[c][c]{\footnotesize{2}}
	\psfrag{002}[c][c]{\footnotesize{3}}
	\psfrag{003}[c][c]{\footnotesize{4}}
	\psfrag{004}[c][c]{\footnotesize{5}}
	\psfrag{005}[c][c]{\footnotesize{6}}
	\psfrag{006}[r][r]{\footnotesize{0}}
	\psfrag{007}[r][r]{\footnotesize{0.2}}
	\psfrag{008}[r][r]{\footnotesize{0.4}}
	\psfrag{009}[r][r]{\footnotesize{0.6}}
	\psfrag{010}[r][r]{\footnotesize{0.8}}
	\psfrag{011}[r][r]{\footnotesize{1}}
	\psfrag{012}[l][l]{\footnotesize{Simple model}}
	\psfrag{013}[l][l]{\footnotesize{Model}}
	\psfrag{014}[l][l]{\footnotesize{Measurement}}
	\psfrag{015}[tc][tc]{\footnotesize{Evaluation video index}}
	\psfrag{016}[bc][bc]{\footnotesize{Consumed energy [J]}}
	\includegraphics[width=0.45\textwidth]{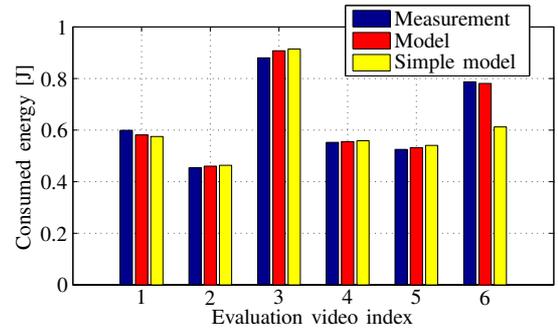}
	\vspace{-0.2cm}
	\caption{Comparison of calculated and measured energy consumption for different evaluation videos. }
	\vspace{-0.2cm}
	\label{fig:cpMeasCalc}
\end{figure}

Figure \ref{fig:cpMeasCalc} shows a comparison of the measured decoding energy, the energy calculated by the accurate model, and the energy calculated by the simplified model. Six exemplary videos are evaluated. The videos are intra coded, 
where coding decisions are made using standard rate-distortion optimization. The video parameters are given in Table \ref{tab:vidParam}. 
The relative error $\varepsilon = \frac{\left| E_\mathrm{dec}-\hat E_\mathrm{dec}\right|}{E_\mathrm{dec}}$ is smaller than $3.2\% $ for the accurate model and, disregarding the low-QP evaluation video 6, smaller than $4.1\%$ for the simplified model. 

\begin{table}
\renewcommand{\arraystretch}{1.3}
\caption{Evaluation video parameters }
\vspace{-0.55cm}
\label{tab:vidParam}
\begin{center}
\begin{tabular}{|r||l||c||r||r|}
\hline
 & Name & Size (pixels) & No. slices & QP \\
\hline\hline
 1 & Basketball Pass & $416\times 240$ & $16$ & $32$\\
\hline
 2 & Basketball Pass & $416\times 240$ & $16$ & $45$\\
\hline
 3 & Race Horse &  $832\times 480$ & $8$ & $45$\\
\hline
 4 & Kimono & $1920\times 1080$ & $1$ & $32$\\
\hline
 5 & Kimono & $1920\times 1080$ & $1$ & $45$\\
\hline
 6 & Bubbles & $416\times 240$ & $8$ & $10$\\
 \hline
\end{tabular}
\end{center}
\vspace{-0.8cm}
\end{table}

\section{Conclusion}
\label{sec:concl}
An accurate as well as a simplified model for the energy consumption of the HEVC intra decoder on mobile platforms has been presented. 
The relative error to the measured consumption has been proven to be lower than $4.1\%$ for the simplified model. The possibility to employ the model in encoders enables novel coding methods aiming at reducing decoding energy consumption. 
Further work is planned to extend the model to inter prediction and loop filters, and to apply it to encoder decisions. 


\section*{Acknowledgment}
This work was financially supported by the Research Training Group 1773 ``Heterogeneous Image Systems'', funded by the German Research
Foundation (DFG).



\bibliographystyle{IEEEtran.bst}
\bibliography{IEEEabrv,D:/Literatur/literatureNeu}
%



\end{document}